\documentstyle[epsf]{npbproc}
% ...
\newcommand{\BI}{Fakult\"at f\"ur Physik, Universit\"at Bielefeld,
        D-4800 Bielefeld 1, Germany.}
\newcommand{\HLRZ}{HLRZ c/o Forschungszentrum J\"ulich,
            P.O.Box 1913, D-5170 J\"ulich, Germany.}
\newcommand{\SCRI}{SCRI, The Florida State University,
                  Tallahassee, FL 32306, USA.}
\newcommand{\TIFR}{Tata Institute of Fundamental Research,
          Homi Bhabha Road, Bombay 400005, India.}

\begin{document}
% title of the article
\title{A STUDY OF SYMMETRY RESTORATION AT FINITE TEMPERATURE\\
  IN THE O(4) MODEL USING ANISOTROPIC LATTICES\thanks%
{Presented by R.~V.~Gavai}}
% use separate \author and \address instructions
% for every author, or group of authors
\author{ R.~V.~Gavai${}^1$,
         U.~M.~Heller${}^2$,
         F.~Karsch${}^{3,4}$,
         B.~Plache${}^4$,
         T.~Neuhaus${}^4$}
\address{${}^1$\TIFR\\
         ${}^2$\SCRI\\
         ${}^3$\HLRZ\\
         ${}^4$\BI}

% a date field is not required
\date{}

% title and author's name as they appear in the running header
\runtitle{O(4) Model at Finite Temperature} %If the title is too long, use
                                                % a shorter title
\runauthor{R. V. Gavai et al.} %Use initials and surnames.  
                     %For 3 or more authors, use
                     %the first author and et al.,
                     %e.g., \runauthor{D. Adams et al.}
% the data below are for the volume and page numbers in the running header.
% Correct numbers will be pasted over the printed lines by the publisher.
\volume{XXX}  % Do not change the field.
\firstpage{1} % Do not change the field.
\lastpage{3}  % Supply the number of pages of your manuscript in the field,
              % e.g., \lastpage{10} for a 10 page manuscript.

\begin{abstract}

Results of investigations of the $O(4)$ spin model at finite temperature
using anisotropic lattices are presented. In both the large $N$ approximation
and the numerical simulations using the Wolff cluster algorithm we find that
the ratio of the symmetry restoration temperature $T_{\rm SR}$ to the Higgs
mass $m_{\rm H}$ is independent of the anisotropy.  We obtain a lower bound
of $0.59 \pm 0.04$ for the ratio, $T_{\rm SR}/m_{\rm H}$, at
$m_{\rm H}a \simeq 0.5$, which is lowered further by
about $10 \%$ at $m_{\rm H}a \simeq 1.$

\end{abstract}

% put the front matter information on paper
\maketitle

% and here comes the text ...

\section{INTRODUCTION}\label{introduction}
Finite temperature investigations of spontaneously broken gauge theories are
of importance to the physics of the very early universe. Two
prime examples are the inflationary universe and the generation of the
baryon asymmetry. Although symmetry restoring phase transitions in
spontaneously broken gauge theories are crucial for these areas, our
knowledge about them comes chiefly from perturbation theory. Motivated by
the desire to learn more about their non-perturbative aspects, lattice
investigations\cite{Rly1,Rly2} of these theories at finite temperature have
been made. Our investigation of the $O(4)$ spin model on anisotropic lattices
is one more step in this direction.  Recall that this model is obtained from
the fundamental $SU(2)$ Higgs-gauge model in its weak gauge coupling limit.
Both the models are expected to be trivial,
giving rise to an upper bound on the Higgs mass in their respective scaling
regions. The finite temperature investigation is aimed at studying the model
in this scaling region to find out about the symmetry restoring phase
transition and to obtain a lower bound on the symmetry restoration temperature
$T_{\rm SR}$. Employing anisotropic lattices has the advantage of being
able to distinguish the finite temperature effects, which could be called
as a special type of finite size effects, from arbitrary finite size effects
since the former have to be independent of the anisotropy in the scaling region.
In addition, anisotropic lattices allow one to study the finite temperature
effects at a correlation length of order unity.

\section{THE ANISOTROPIC O(N) MODEL}\label{the models}
On anisotropic lattices $L^3 \times \xi L_t$, the $O(N)$ spin model is defined
by the action
\begin{eqnarray}
S = - N \beta (\gamma \sum_x S_x \cdot S_{x + \hat 0} +
{1 \over \gamma} \sum_{x,j} S_x \cdot S_{x + \hat j} )
\end{eqnarray}
where $S_x \in O(N)$, $\forall x$, and $\beta$(or $\kappa =  N \beta / 2$
for $O(4)$) is the coupling on the isotropic lattices
for which the anisotropy coupling $\gamma$ is unity. The physical volume and
temperature are respectively given by $V=L^3a^3$ and $T=1/\xi L_t a_t$. Since
$\xi=a/a_t$, varying $\xi$ on the lattices above amounts to holding the
temperature constant in units of $a^{-1}$,
apart from possible quantum renormalization effects.

In both the large $N$ approximation and the numerical simulations our procedure
to investigate finite temperature effects was the following.  For a given value
of the anisotropy coupling $\gamma$, we obtained the critical coupling on an
$L^3 \times \xi L_t$ lattice by setting $\gamma$ to its classical value $\xi$.
Our results justify this choice {\it a posteriori}.  $\beta_c$ in the large $N$
limit is obtained by solving numerically the saddle point equation
\begin{eqnarray}
\beta_c =  {1 \over L_t L^3} \sum_p {}' {1 \over D(p) }
\end{eqnarray}
for $L \rightarrow \infty$,  where $D(p)$ is given by
\begin{eqnarray}
D(p) = 4 \xi^2 \sin^2 (\half p_0) +
4 \sum_j \sin^2 (\half p_j) ~~,~~
\end{eqnarray}
with the momenta $p_\mu$ given by $p_\mu = 2 \pi n_\mu / N_\mu$,
$n_\mu = 0, \dots , N_\mu - 1$, where $N_0 = \xi L_t$ and $N_j = L$.
The prime on the sum in Eq. (2)
indicates that the zero mode, $p = 0$, is being left out.
In Monte Carlo(MC) simulations the unique crossing point of the cumulant
$g_{\rm R}= \langle M^4 \rangle/\langle M^2 \rangle^2$ for various volumes
$L^3$ yields $\kappa_c(\infty, \xi L_t)$.
Here $M$ is the order parameter, defined by $M=\langle | \sum_x S_x |
\rangle /\xi L^3 L_t$.  Alternatively, one may use the peak position of the
susceptibility, $\chi=\langle M^2 \rangle - \langle M \rangle^2$, to define
$\kappa_c(L,\xi L_t)$. Using the critical exponents of the $O(4)$ model in
three dimensions, $\kappa_c(\infty, \xi L_t)$ can then be obtained using the
finite size scaling theory.

The Higgs mass at zero temperature was then determined at the
$\kappa_c(\infty, \xi L_t)$ by studying the zero momentum connected
correlation functions of the spin variables on $L^3 \times \xi L$ lattices in
both the spatial and the temporal directions.  From the exponential fall-off of
the correlation functions, the Higgs mass, $m_{\rm H}a$, can be obtained using
standard methods, such as fits or local distance dependent masses.
Demanding Euclidean invariance and by appropriately scaling the temporal
direction to match these correlation functions, we determined corrections to
the relation $\xi=\gamma$. They were found to be  5-10\% for all
$\gamma$-values we studied, indicating that the quantum corrections to the
anisotropy are small. The same conclusion was also obtained in the large $N$
limit in the symmetric phase.

\section{RESULTS}\label{results}

%Figure 1
\begin{figure}
\vspace{3.0in}
\includegraphics{on_fig1.psc}
%\epsfxsize=\columnwidth
%\epsffile{on_fig1.psc}
\caption{ $g_{\rm R}$ and $\chi$ as a function of $\kappa$ for $\xi=1.5$
on $L^3 \times 6$ lattices for $L=18$ and 24.}
\end{figure}
%Figure 1

Fig. 1 exhibits our results for both $g_{\rm R}$
and $\chi$ on $18^3 \times 6$ and
$24^3 \times 6$ lattices for $\xi =1.5$.   We used the spectral density
method to obtain the smooth curves shown from our data, shown by crosses.
Similar results have also been obtained for other values of $\xi$ and $L_t$.
We have used $\xi=1$, 1.5 and 2, $L_t=2$, 4, 6 and $L=18$ and 24.
In each case we obtained $\kappa_c(\infty, \xi L_t)$ by using both the
crossing point of $g_{\rm R}$ and the finite size scaling of the peak position
of the susceptibility.  Both estimates were always found to be consistent,
although we preferred to use the former for determining $m_{\rm H}$.

%Table 1
\begin{table}
\begin{tabular}{|c|c|c|c|}                 \hline
$~~~~~~\xi$&    1            &    1.5          &    2            \\
$L_t~~~~$  &                 &                 &                 \\ \hline
     2     &    -            &    -            & 0.54 $\pm$ 0.04 \\ \hline
     4     & 0.59 $\pm$ 0.04 & 0.58 $\pm$ 0.03 & 0.59 $\pm$ 0.03 \\ \hline
     6     & 0.80 $\pm$ 0.01 &    -            &    -            \\ \hline
\end{tabular}
\caption{$T_{\rm SR}/m_{\rm H}$ as a function of $\xi$ and $L_t$.}
\end{table}
%Table 1

At each coupling, the Higgs mass $m_{\rm H}$ was obtained from the plateau in
the local distance-dependent masses, defined as $\ln
(C(t)/C(t+1))$, where $C(t)$ is the zero momentum correlation function. Again
we checked that a fit to the data of an exponential form yielded
consistent results with these estimates.
Using these results, the ratio $T_{\rm SR}/m_{\rm H} = (L_t m_{\rm H}a)^{-1}$
shown in Table 1 is obtained for various $\xi$ and $L_t$.
The $\xi$-independence of the ratio is obvious.
Recall that $m_{\rm H} \rightarrow 0$, as one approaches $\kappa_c$, i.e., as
$L_t$ grows. Thus, depending on the choice of value of the correlation length
up to which an effective theory can be defined, one obtains a lower bound
on the ratio $T_{\rm SR}/m_{\rm H}$.  From Table 1, one sees this bound to be
$0.59 \pm 0.04$ for a correlation length of $\sim$ 2, which decreases
by 10\% for $m_{\rm H}a \simeq 1.$

%Figure 2
\begin{figure}
\vspace{2.9in}
\includegraphics{on_fig2.psc}
%\epsfxsize=\columnwidth
%\epsffile{on_fig2.eps}
\caption{ Large $N$ results for $T_{\rm SR}/m_{\rm H}$ and
$T_{\rm SR}/v_{\rm R}$ as a function of $\xi$ for various $L_t$.}
\end{figure}
%Figure 2

Considering the fluctuations around the saddle point of the large $N$ limit,
one can obtain the Higgs mass $m_{\rm H}$ at $\beta_c(\xi L_t)$, while the
corresponding renormalized vacuum expectation value of the field is given by
$v_{\rm R}^2 = \beta_c(\xi L_t) - \beta_c(\infty)$.  Fig. 2 shows these large
$N$ results for $T_{\rm SR}/m_{\rm H}$ and $T_{\rm SR}/ v_{\rm R}$.
Both are seen to be clearly independent of $\xi$. Further, the latter seems
to be independent of $L_t$ for $L_t \ge 4$.  Qualitatively, the large $N$
limit seems to reproduce all the features of the Monte Carlo(MC) data well.
However, quantitatively, the large $N$ results seem to lie systematically
lower than the MC results by $\sim$15 \%.  It would be interesting to
check whether the early scaling evident in the MC data for $L_t=2$ is
real by simulating the theory at more $\xi$ values and also by studying
the $T_{\rm SR}/v_{\rm R}$ ratio.

\acknowledge{
One of us (R.V.G.) is thankful to the organizers of ``LATTICE 91''
for their financial support which enabled him to present this work.}

% the bibliography comes at the end


\begin{thebibliography}{9}
\bibitem{Rly1}
H. G. Evertz, J. Jers\'ak and K. Kanaya, Nucl. Phys.
{\bf B285[FS19]} (1987) 229;
P. H. Damgaard and U. M. Heller, Nucl. Phys. {\bf B304} (1988) 63.
\bibitem{Rly2}
U. M. Heller, Phys. Lett. {\bf B191} (1987) 109;
K. Jansen and P. Seuferling, Nucl. Phys. {\bf B343} (1990) 507.
\end{thebibliography}
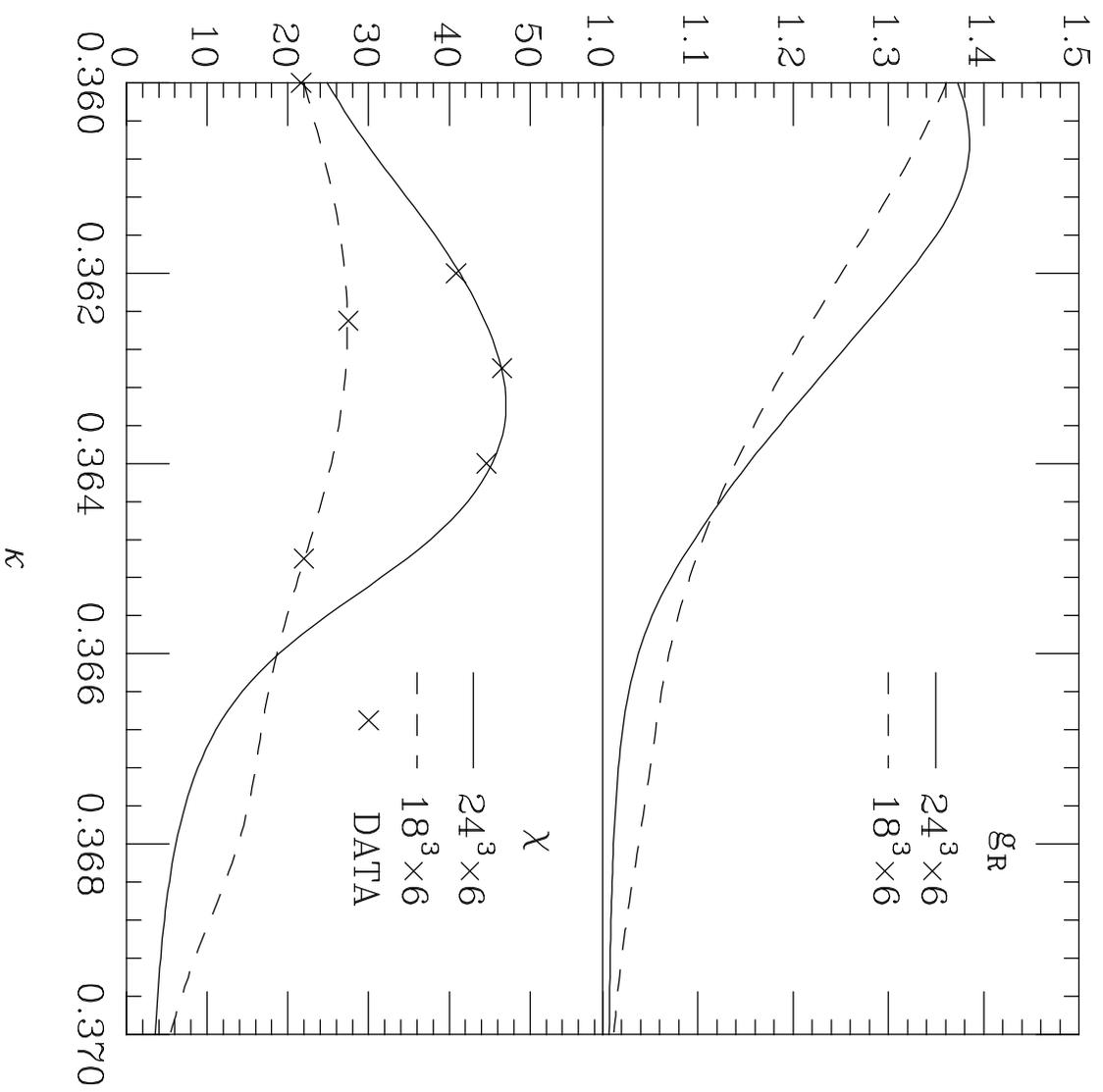
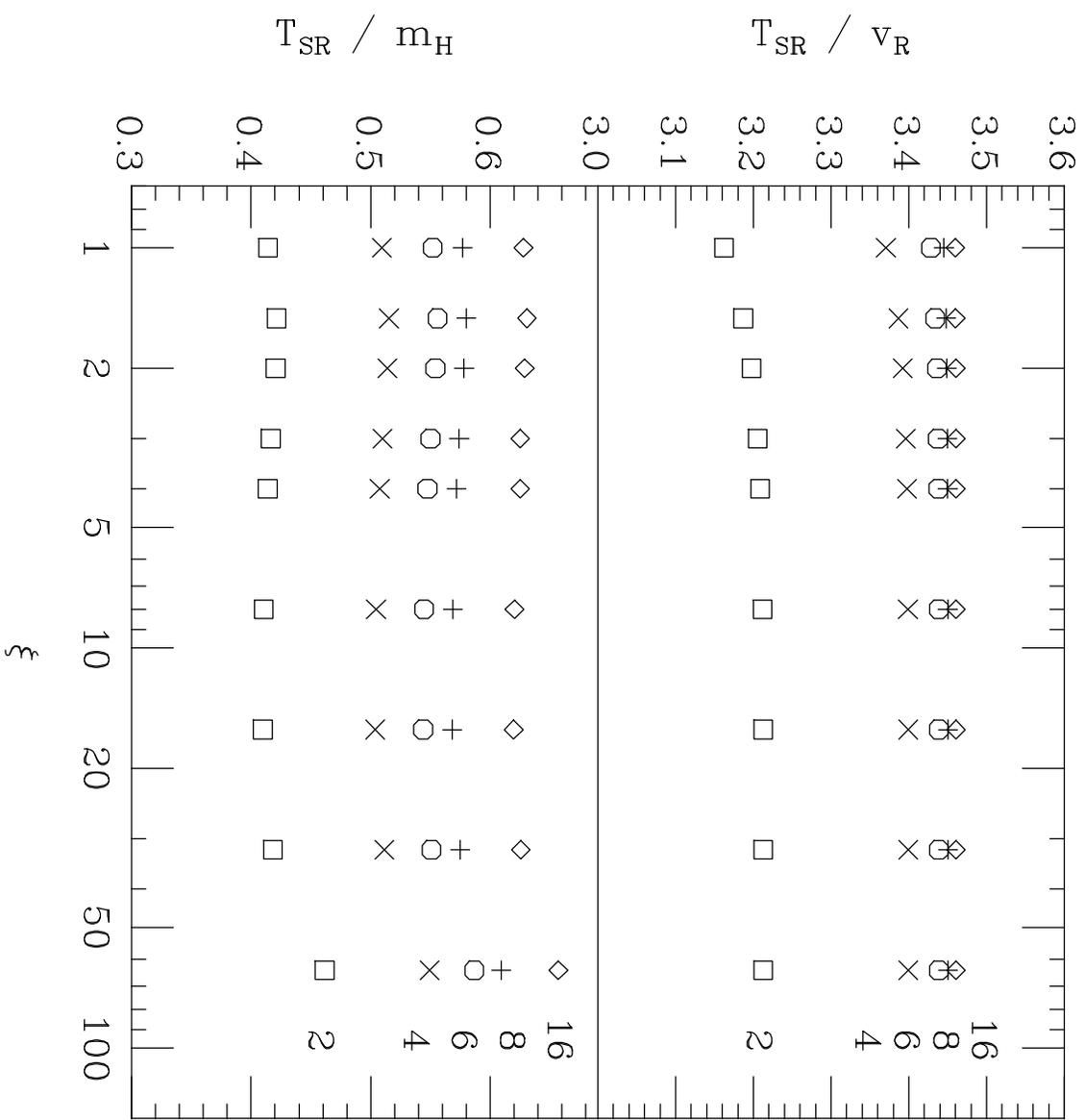
\end{document}